\documentclass[twocolumn,amsmath,amssymb,10pt,nofootinbib,superscriptaddress]{revtex4}
\usepackage{ graphicx, float,amsmath, amsmath, amssymb}
\usepackage{xcolor} 
\usepackage[export]{adjustbox}
\usepackage[breaklinks, colorlinks, citecolor=blue]{hyperref}
\usepackage[labelsep=period]{caption}
\usepackage[normalem]{ulem}
\usepackage{mathtools}
\usepackage{siunitx}
\usepackage{soul}
\usepackage{physics}
\usepackage{minitoc}

\usepackage{bm}
\usepackage{xcolor}

\definecolor{alizarin}{rgb}{0.82, 0.1, 0.26}

\def\be{\begin{equation}}
\def\ee{\end{equation}}
\def\bea{\begin{eqnarray}}
\def\eea{\end{eqnarray}}
\def\bse{\begin{subequations}}
\def\ese{\end{subequations}}

\begin{document}
\setlength{\parindent}{0cm}

\title{Stochastic background of gravitational waves from cosmic-rays}

\author{Aur\'{e}lien Barrau}%
\affiliation{%
Laboratoire de Physique Subatomique et de Cosmologie, Universit\'e Grenoble-Alpes, CNRS/IN2P3\\
53, avenue des Martyrs, 38026 Grenoble cedex, France
}

\author{Juan Garc\'ia-Bellido}
\affiliation{Instituto de F\'isica Te\'orica UAM/CSIC, Universidad Aut\'onoma de Madrid, Cantoblanco 28049 Madrid, Spain}

\author{Killian Martineau}%
\affiliation{%
Laboratoire de Physique Subatomique et de Cosmologie, Universit\'e Grenoble-Alpes, CNRS/IN2P3\\
53, avenue des Martyrs, 38026 Grenoble cedex, France
}

\author{Daryna Yushchenko}
\affiliation{%
Laboratoire de Physique Subatomique et de Cosmologie, Universit\'e Grenoble-Alpes, CNRS/IN2P3\\
53, avenue des Martyrs, 38026 Grenoble cedex, France
} 
\affiliation{Princeton University, Princeton, NJ, 08544, USA}



\date{\today}
\begin{abstract} 
Cosmic-rays are charged particles moving in magnetic fields. They not only emit well-known synchrotron photons but also gravitational radiation. We clarify the characteristics of the gravitational wave signal in this specific situation and underline some unexpected features. A phenomenological approximation for the radiated power is given. We derive the shape and peaking frequency of the associated stochastic backgrounds of gravitational waves for both electrons and protons, either of galactic or extra-galactic origin.
\end{abstract}
\maketitle

\section{Introduction}

The status of gravitational waves (GWs) has generated a long and fascinating controversy (see \cite{Gomes:2023xda} and references therein for a recent review). Most issues are now solved and GWs constitute one of the clearer and stronger predictions of general relativity (GR). They have been experimentally detected in the nHz
(see \cite{NANOGrav:2023gor}) and kHz bands (see \cite{LIGOScientific:2020ibl,LIGOScientific:2021djp}). The LISA project \cite{Colpi:2024xhw} will open the mHz window, and the Einstein Telescope \cite{Badaracco:2024kpm} should drastically improve the sensitivity of terrestrial antennas. Measurements at even higher frequencies are also being seriously considered 
(see \cite{Aggarwal:2020olq} and references therein), in particular thanks to electromagnetic resonant or superconducting cavities \cite{Berlin:2021txa, Berlin:2023grv}, but also through LASER beams \cite{Vacalis:2023gdz}.\\

Among the expected sources of GWs, there are coherent signals (e.g. binary systems of black holes, neutron stars, exotic compact objects, etc.), stochastic backgrounds (e.g. GWs from the inflationnary era, GWs generated by the cosmological microwave background, GWs from phase transition, etc.), and bursts (e.g. GWs from superradiance and hyperbolic encounters). This study aims to investigate a possible new background due to cosmic rays spiraling in galactic magnetic fields. Let us make it clear from the start that the expected order of magnitude of the associated power is obviously small -- this can be hypothesized before any explicit calculation. However, our goal is {\it not} to suggest a forgotten measurable signal. In fact, 
we believe that the theoretical framework is precisely reliable enough to allow a computation which, in itself, reveals an interesting and non-trivial phenomenon of our Universe, independent of any experimental detection. To the best of our knowledge, this GW background has never been investigated and its main characteristics are not easy to infer, hence this study's relevance. The article is written in a very pedagogical way so that the non-specialist reader gets clear ideas about the situation, and the opportunity is taken to clarify a few confusing points as some results are highly non-intuitive.\\

We first establish the strain and radiated gravitational power for a charged particle in a magnetic field, both in the non-relativistic and relativistic regimes. 
In both cases the particle trajectory is assumed to be completely determined by the electromagnetic interaction. The result is compared with electromagnetic synchrotron radiation and an approximated formula valid in all regimes is derived. The interstellar cosmic-ray spectra are then written as functions of the appropriate (and unusual) variables. The backgrounds of both galactic and extra-galactic GWs for electrons and protons are then computed. 

\section{Strain and power: non-relativistic limit}

It is straightforward to calculate the strain generated by a particle in a circular orbit. The equations of motion read $\ddot{x} = \omega_B \dot{y}, \ddot{y} = \omega_B \dot{x}$ for an orbital pulsation $\omega_B$. This allows us to trivially calculate the second mass moments defined as $M^{ij}(t) = \int d^3x \rho(\textbf{x},t)x^i x^j$. The usual quadrupole formula \cite{Maggiore:2007ulw} -- which works surprisingly well \cite{Blanchet:2019zlt} -- leads to the two polarisations\footnote{This is quite nontrivial as the system is not isolated.}:
\begin{eqnarray}
h_+(t) &=& \frac{4m R^2\omega_B^2}{r}\frac{G}{c^4} \left(\frac{1 + \cos^2(\theta)}{2}\right)\cos(2\omega_Bt),\\
h_{\times}(t) &=& \frac{4 m R^2\omega_B^2}{r}\frac{G}{c^4} \cos(\theta) \sin(2\omega_B t),
\end{eqnarray}

where $\theta$ is the angle between the axis of orbit and an observer located at a distance $r$ from the source, $m$ is the mass of the particle, and $R$ is the radius of the trajectory. It should be noticed that at a given angular frequency, the strain increases as a function of the distance between the masses. Although perfectly correct and intuitive, this behaviour is to be contrasted with what happens for self-gravitating systems where the strain increases with decreasing distances between the two bodies (as the dynamics is obviously {\it not} taking place at fixed angular frequency in this case). Maybe surprisingly, the strain depends {\it only} on the speed of considered particle.\\

We now focus on the maximum value of the strain and consider the case of a charged particle in a magnetic field $B$ 
with gyro-radius $R=\gamma m V / (qB)$ and pulsation $\omega_B=qB/(\gamma m)$ such that V is the linear speed and $\gamma$ is the Lorentz factor. Although in this section the calculation of the emitted GWs 
is performed with the assumption $V\ll c$, we still keep the $\gamma$ factor when it allows a better definition of the orbit. This case simplifies the strain to 
\begin{equation}
    h = \frac{4 G m}{r c^2} \beta ^2,
    \label{strain}
\end{equation}
with $\beta=V/c$. Quite interestingly, the generated strain is independent 
of the magnetic field. This is mathematically obvious when is it kept in mind that, as stated  above, it  depends only upon the speed, with the latter being unaffected by the magnetic field. Still, it might seem inconsistent that for values of $B$ arbitrarily small the strain remains non-vanishing and keeps the very same value than for large magnetic fields. How could a particle moving nearly in straight line generate the same strain than a particle following a highly curved path? The reason is simple: the associated energy is not the same. As the orbital frequency decreases with smaller magnetic fields, so does the energy carried out by the emitted GWs (whose frequency is twice the orbital one, $f_{GW} = 4\pi \omega_B$). One then hopefully recovers that a particle moving at constant speed in straight line does not generate GWs. Still, this means that a charged particle moving in a stochastic -- both in direction and amplitude -- magnetic field generates a constant strain.\\

Using the classical expression of the power per unit solid angle as a function of the strain in the TT gauge,
\begin{equation}
    \frac{dP_{NR}}{d\Omega}=\frac{r^2c^3}{32\pi G}<\dot{h}_{ij}^{TT}\dot{h}_{ij}^{TT}>,
\end{equation}
one is led to \cite{Maggiore:2007ulw}:
\begin{equation}
    P_{NR}=\frac{32}{5}\frac{Gm^2}{c^5}R^4\omega_B^6=\frac{32}{5}c G m^2 \frac{\beta^6}{R^2},
    \label{nonrelat}
\end{equation}
in the non-relativistic ($NR$) case. 
Expressing $P_{NR}$ in the variables of interest leads to
\begin{equation}
    P_{NR}=\frac{32}{5}\frac{G q^2 B^2}{c} \beta^4 (1-\beta^2),
    \label{NR2}
\end{equation}

where the $1 - \beta^2$ term is due to the relativistic correction of the orbit. This, obviously, can be written as a function of energy but the (normalized) speed is, in this case, the more natural variable.

\section{Relativistic case}

The relativistic case, $\beta\sim 1$, is not as straightforward. The very existence of ``gravitational synchrotron radiation" has long been debated in the early days of GW physics. A rigorous treatment was finally given in \cite{pust}. The main ideas of this work are summarized below but we refer any interested reader to the original paper for more details. In this regime the emission of gravitational waves comes from two distinct origins: the usual mass tensor of the particle but also the electromagnetic stresses caused by the existence of the non-vanishing charge. Interestingly, those two contributions are of the same order of magnitude. 
The evolution of the spacetime perturbation thus follows the wave equation
\begin{equation}
    \Box \bar{h}^{\mu \nu} = - \frac{16 \pi G}{c^4} \left( T_{P}^{\mu \nu} + T_{EM}^{\mu \nu} \right), 
\end{equation}
with $\bar{h}_{\mu \nu}=h_{\mu \nu} - \frac{1}{2} \eta_{\mu \nu} h^{\alpha}_{\alpha}$, $\eta_{\mu \nu}$ being the Minkowski metric, and $T_{P}$ and $T_{EM}$ being respectively the particle and electromagnetic field stress-energy tensors. They can be expressed as

\begin{equation}
    T_{P}^{\mu \nu} = m c^2 u^\mu u^\nu \sqrt{1 - \beta^2} \delta \left( \vec{x} - \vec{R} \right),
\end{equation}
$u^\mu$ being the 4-velocity, $\vec{R}(t)$ the particle trajectory, and

\begin{equation}
    T_{EM}^{\mu \nu} = \frac{1}{\mu_0}  \left(F^{\mu \lambda}F^{\nu}_{\lambda} - \frac{1}{4} \eta^{\mu \nu} F_{\alpha \beta} F^{\alpha \beta}\right),
\end{equation}
with $\mu_0$ the vacuum permeability and $F^{\mu \nu}$ the electromagnetic field tensor. The total electromagnetic field entering the above formula is the sum of both the particle field $F^{\mu \nu}_P$ and the external field $F^{\mu \nu}_0$. The electromagnetic stress-energy tensor being quadratic in the fields, one approximates

\begin{equation}
    T_{EM} \sim F_0 F_0 + 2 F_0 F_P + F_P F_P.
\end{equation}
The first term is time-independent and does not generate any gravitational radiation. The last term corresponds to the particle electromagnetic self-energy and is discarded after renormalization. The only remaining term contributing to the gravitational radiation is thus the cross-term $2 F_0 F_P$. Going into Fourier space, the metric perturbation is written as :

\begin{equation}
    \bar{h}^{\mu \nu}(\omega, \vec{k}) = - \frac{4 G i}{\pi c^4 r} \int_{- \infty}^{+\infty} \frac{T^{\mu \nu}_{P}(\omega,\vec{k})+T^{\mu \nu}_{EM}(\omega,\vec{k})}{k^2 - \omega^2/c^2} e^{-i \vec{k}.\vec{r}} k dk,
\end{equation}
where $\vec{r} = \vec{x}-\vec{R}$. Specifying the expressions of $T^{\mu \nu}_{P}$ and $T^{\mu \nu}_{EM}$ as in \cite{pust}, and solving using the residues leads to

\begin{equation}
    \bar{h}^{\mu \nu}(\omega, \vec{k}) = \frac{4 G m}{c^2 R \sqrt{1 - \beta^2}} e^{i \delta} I^{\mu \nu}(\omega, \vec{k}),
    \label{h Gertsenstein}
\end{equation}
where the detailed expressions of the phase $\delta$ and of the $I^{\mu \nu}$ functions are given in \cite{pust}.

The power of gravitational radiation writes

\begin{equation}
    \frac{dE}{dt} = \frac{c^3 r^2}{32 \pi G} \int \langle \dot{h}_{ij}^{TT} \dot{h}^{ij,TT} \rangle  d\Omega,
    \label{gravitational power}
\end{equation}
where the index $TT$ refers to quantities evaluated in the transverse-traceless gauge (in which $\bar{h}^{\mu \nu} = h^{\mu \nu}$), $d\Omega$ is the differential solid angle and $< >$ denotes an averaging over the full revolution time of the particle. Plugging Eq. (\ref{h Gertsenstein}) into Eq. (\ref{gravitational power}) and neglecting terms that are small in the non-relativistic regime leads to the following expression for the ultra-relativistic ($UR$) radiated power \cite{pust}:
\begin{equation}
    P_{UR} = \frac{39}{8} c G m^2\frac{\gamma^4}{R^2}.
    \label{ultra}
\end{equation}

To allow an easy comparison with the non-relativistic case, one can express $R$ and $\gamma$ as functions of the particle velocity:
\begin{equation}
P_{UR} (\beta) = \frac{39}{8} \frac{Gq^2 B^2}{c} \frac{1}{\beta^2(1-\beta^2)}.
\label{URpower}
\end{equation}
As expected, this result is drastically different from Eq. (\ref{NR2}).
Several additional considerations on the structure of synchrotron gravitational radiation are given in \cite{Davis:1972dm}. In particular it is stressed that the complexity of the tensor field does not allow for a straightforward extrapolation from the usual electromagnetic synchrotron radiation.
Equation (\ref{URpower}) is to be contrasted with the one used in \cite{DiambriniPalazzi:1987uy} to investigate the GW generation in storage rings of accelerators, where the $\gamma$ dependence is incorrect. When comparing different expressions one should be careful at identifying which quantities are fixed. In our case, the magnetic field is fixed and the radius of curvature obviously depends on the energy, or speed, of the considered particle. The emitted gravitational power scales as $\gamma^4$. When considering a particle in a storage ring however, the radius is fixed and the gravitational power is proportional to $\gamma^2$. The non-monotonic behavior of Eq. (\ref{URpower}) might seem strange but is entirely an artifact as this expression is valid only in the ultra-relativistic limit. As we shall show later, the correct expression always features a positive derivative.\\

An expression for the spectrum of the generated GWs has been given in \cite{Chen:2021rvg}:
\begin{eqnarray}
    \frac{dP_{UR}}{dx}&=&\frac{3\sqrt{\pi}}{32}\frac{Gm^2\gamma^4\omega_B^2}{c}\\ \nonumber
    &\times&\left( 3x^{-\frac{1}{3}}\Phi(y)-5x^{\frac{1}{3}}\Phi'(y)+3x\Phi_2(y) \right),
    \label{GW power spectrum}
\end{eqnarray}
such that $\omega_c=\gamma^3\omega_B$, $x=\omega/\omega_c$, $y=x^{2/3}$, $\Phi$ the Airy function, and 
\begin{equation}
\Phi_2(y)=\frac{y^{\frac{1}{2}}}{2^{\frac{2}{3}}\pi^{\frac{1}{2}}}\int_{-\infty}^{+\infty}\Phi^2(y(1+z^2)/2^{\frac{1}{3}})dz.
\end{equation}
The critical frequency $\omega_c$ corresponds to the peak of the emitted synchrotron electromagnetic radiation. This basically follows from writing
\begin{equation}
    P_{UR}=\frac{r^2}{32\pi G}\int d\Omega \left( \partial_0\bar{h}^{\mu\nu}\partial_i\bar{h}_{\mu\nu}-\frac{1}{2}\partial_0\bar{h}^{\mu}_{\mu}\partial_i\bar{h}^{\nu}_{\nu}\right) n^i,
\end{equation}
where $\hat{n}$ is a unit vector, re-expressing the integrand with the stress-energy tensor and harmonically expanding the later.

It can easily be seen from this expression -- and as argued previously in \cite{Khalilov:1972xau,Aliev:1980du,Chen:2021rvg} -- that this spectrum is monotonically decreasing with frequency and 
thus dominated by the fundamental mode. 
We shall then assume, as a good enough approximation for this work, that the GWs are always emitted at $f_{GW}=4\pi \omega_B$. This is to be contrasted with the spectrum of electromagnetic radiation which has a maximum at $f_{\gamma}=2\pi \gamma ^3 \omega_B$. This difference in the frequency spectra between electromagnetic and gravitational synchrotron radiations has also been found for relativistic particles orbiting in Schwarzschild and Kerr geometries \cite{Davis:1972dm,Chitre:1972fv,PhysRevD.7.1002,Ternov:1975ut}. \\

It is clear that the power given by Eq. (\ref{ultra}) is enhanced by a factor $\gamma^4$ at high energy when compared to Eq.(\ref{nonrelat}) (which itself is 
suppressed by a factor $\beta^6$ at low energy). As the spectra of cosmic-rays (CRs) decrease with energy, it is not {\it a priori} obvious to guess which CR energy contributes dominantly and the associated frequency. As a useful phenomenological approximation valid at all energies, we now use:

\begin{equation}
P \approx \zeta Gc \frac{m^2 \beta^6 \gamma^4}{R^2},
\label{power}
\end{equation}

with $\zeta$ of order unity (the best match is obtained for $\zeta\approx 5$). 
Without any subscript, $P$ refers to the GW power expressed in a way that is (approximately) valid without restriction. One immediately sees that the ultra-relativistic behavior is, as expected, scaling with $\gamma^4$ whereas the low-energy limit is, also correctly, in $\beta^6$. This combined expression is consistent with both the non-relativistic and ultra-relativistic cases. In practice, Eq. (\ref{power}) works more than adequately at the level of precision required in this study.\\

For the sake of completeness, using the explicit form of $R$ for the considered trajectories, 
Eq. (\ref{power}) can be re-expressed as a function of velocity,
\begin{equation}
    P (\beta)=\frac{\zeta G q^2 B^2}{c} \frac{\beta^4}{1-\beta^2},
\end{equation}
or, alternatively, in terms of (total) energy:
\begin{equation}
P (E) = \zeta G q^2 B^2 \frac{{\left(E^2-m^2c^4\right)}^2}{m^2c^5E^2}.
\end{equation}

Although expressing the power in terms of the momentum  
leads to a simpler expression, energy is, in this case, a more relevant variable since the relativistic generalization of the Larmor pulsation does depend on the energy of the particle spiralling in a magnetic field (this was not the case in the non-relativistic regime). Expressing the power in terms of the energy therefore allows us to take advantage of the one-to-one correspondence between the energy of the particle and its orbital frequency, hence with the GW frequency, $f_{GW} = 2f_B= \frac{qBc^2}{\pi E}$.

Should the considered particles be non-relativistic, they would emit nearly monochromatic GWs, whatever their speed. However, when considering cosmic-rays, the spectrum of GWs might become very extended due to very high energy cosmic particles. As it will be useful in the following, we now re-express the GW power as a function of the frequency:


\begin{equation}
P (f) = \zeta G q^2 B^2 \frac{{\left(\left(\frac{qBc^2}{\pi f}\right)^2-m^2c^4\right)}^2}{m^2c^5{\left(\frac{qBc^2}{\pi f}\right)^2}}.
\label{powerfreq}
\end{equation}

\begin{figure}
    \centering
\includegraphics[width=0.5\textwidth]{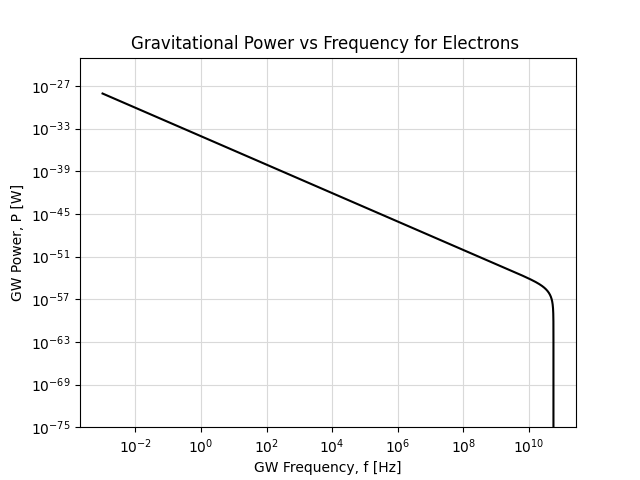}
    \caption{Power lost as GWs by an electron in a magnetic field of 1 T, as a function of the frequency of the GWs (that is twice the orbital frequency of the electron).}%
    \label{fig1}
\end{figure}

The frequency is the variable of interest for this study as we aim at determining the frequency spectrum of gravitational waves.
The GW frequency $f$ 
varies, in principle, between $0$, which corresponds to $E\rightarrow \infty$, and $qB/(\pi m)$, which corresponds to a particle at rest. Maybe surprisingly (although it is quite obvious), large frequencies correspond to small energies. When all other parameters are the same, the energy of a GW increases with its frequency but Eq. (\ref{powerfreq}) is -- as seen in Fig. \ref{fig1} -- a monotonically decreasing function of $f$. This is possible because the higher particle energies, associated with lower frequencies, largely compensate for this intrinsic property of GWs. It is 
crucial to keep in mind that the magnetic field remains fixed and the frequency variation is entirely due to the energy of the considered particles.

\section{Comparison with electromagnetic synchrotron radiation}

The power carried out by the electromagnetic radiation is given by the relativistic Larmor formula:

\begin{equation}
P_\gamma = \frac{q^2c}{6 \pi \epsilon_0} \frac{\beta^4 \gamma^4}{R^2}.
\end{equation}

This immediately leads to the ratio:

\begin{equation}
\frac{P}{P_\gamma} = {6 \pi \epsilon_0 G \zeta} \left( \frac{m}{q} \right)^2\beta^2.
\end{equation}

The GW power is strongly suppressed (by the $\beta^2$ term), when compared to the electromagnetic synchrotron power in the low-energy limit. However, both are scaling as $\gamma^4$ (for a given radius $R$) at high energy. Maybe less obviously, this causes the ratio 
to be constant -- even if $R$ does vary with the energy -- in the relativistic limit. Interestingly, once again, the value of the magnetic field does not play any role. \\

For typical values of the parameters, the electromagnetic synchrotron radiation strongly dominates, as expected. However, quite impressively, for a unit charge, the mass above which the GW power dominates (at relativistic speeds) is $m_c=6.8\times 10^{-10}$ kg. This is 2 orders of magnitude below the Planck mass and therefore not absurd for an elementary particle. Although this is not directly related to this work, it is worth stressing that in principle 
the power carried away by gravitational radiation may therefore become higher than the power carried by electromagnetic radiation for hypothetical ultra-heavy relativistic particles.

\section{Galactic cosmic-ray spectra}

Because of solar modulation, the interstellar flux of cosmic-rays cannot be directly determined by measurements. Recently, impressive progress has been made using the Voyager 1 observations performed beyond the heliopause by AMS, BESS, and PAMELA data in Earth orbit \cite{Ghelfi:2015tvu,Bisschoff:2019lne}. We use the parameterizations obtained with the GLAPROP code -- reproducing the electron, proton, Helium and Carbon observations -- to estimate the interstellar fluxes from data. It should however be noticed that, for protons, we use a parametrization that differs from the one given in \cite{Bisschoff:2019lne} as it better fits the interstellar fluxes -- at least in the energy range relevant for this study.\\

The interstellar electron flux reads
\begin{eqnarray}
 \Phi_{\rm elec}(E_k) &=& \frac{255.0}{\beta ^2} \, \left(\frac{E_k}{E_{k0}}\right) ^{-1} \left(\frac{E_k/E_{k0} +0.63}{1.63} \right) ^{-2.43}\notag\\ 
 &+& 6.4 \, \left(\frac{E_k}{E_{k0}}\right) ^{2} \left(\frac{E_k/E_{k0} + 15.0}{16.0}\right) ^{-26.0},
\label{expr1}
\end{eqnarray}

whereas the proton flux is given by
\begin{eqnarray}
\label{expr3.1}
\Phi_{\rm p}(E_k) &=& 200.0\left(\frac{E_k}{E_{k0}}\right)^{0.1} \left(\frac{E_{k}/E_{k0}+0.7^{0.98}}{3+0.7^{0.98}}\right)^{-3.2} \notag\\
&+& 30.0\left(\frac{E_k}{E_{k0}}\right)^{2.0}\left(\frac{E_k/E_{k0}+8.0}{9.0}\right)^{-12.0}.
\end{eqnarray}

Both are given in part.m$^{-2}$.s$^{-1}$.sr$^{-1}$.GeV$^{-1}$ as a function of the kinetic energy $E_k=E-mc^2$ (in GeV), with $E_{k0}$ = 1\,GeV.\\

\begin{figure}
    \centering
\includegraphics[width=0.5\textwidth]{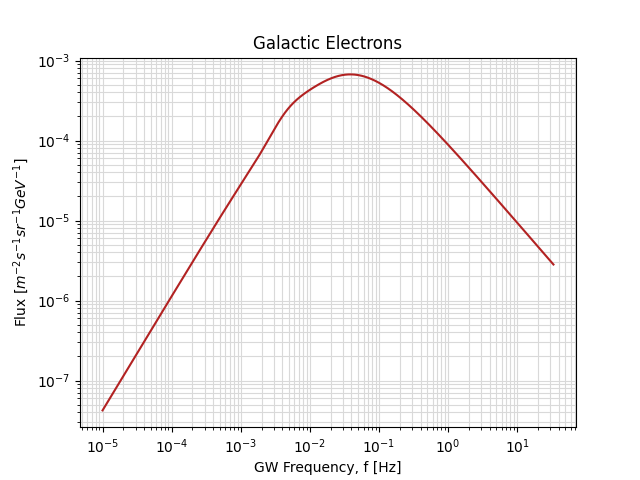}
    \caption{Interstellar spectrum of cosmic electrons as a function of the frequency of the emitted GWs.}%
    \label{fig2}
\end{figure}

\begin{figure}
    \centering
\includegraphics[width=0.5\textwidth]{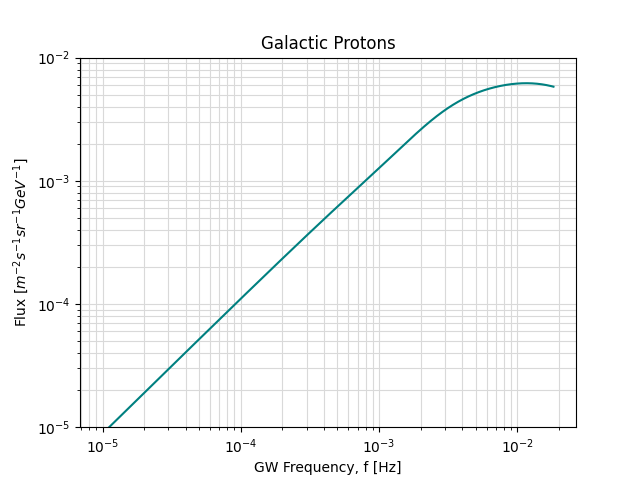}
    \caption{Interstellar spectrum of cosmic protons as a function of the frequency of the emitted GWs.}%
    \label{fig3}
\end{figure}

In the following we assume that the galactic magnetic flux is roughly (spatially) constant in amplitude with $B_{\text{gal}}\approx 6 \mu$G \cite{Beck:2007}. It is not uniform across directions, hence the random walk of cosmic-rays. This is why we assume the emission to be, in average, isotropic at each space point. We shall take advantage of the one-to-one correspondence between the energy of the particles and their orbital frequency $f_B$ (which is one half of the emitted GW frequency $f$) to rewrite the spectrum of cosmic-rays per unit surface, time, frequency and solid angle as a function of frequency $\frac{d^4N}{dSdtdfd\Omega}(f)$. This is easily done by replacing the kinetic energy in the previous parametrization by its expression as a function of the frequency, by changing the units to appropriate ones and by noticing that
\begin{equation}
    df=\frac{qBc^2}{2\pi}f^2dE,
\end{equation}
which implies an additional $(qB/2\pi)(c/f)^2$ factor. We however do no need the flux of cosmic-rays received on a given surface but the volume density of cosmic rays per unit frequency.
After integrating over the solid angle and dividing by the speed associated with the considered kinetic energy (or frequency), one is led with the volume density of CRs per frequency $\frac{d^2N}{dVdf}(f)$.\\

The resulting spectrum for electrons is displayed in Fig. \ref{fig2} whereas protons are shown in Fig. \ref{fig3}. This unusual choice of variable is 
suited for our purpose but requires some thinking: the differential spectrum of particles is not plotted here as a function of their intrinsic properties but as a function of the frequency of the GWs they emit. 
It should be, once more, noticed that the left side of the plots, that is low frequencies, correspond to high energies, whereas the right side corresponds to low energies. The maximum frequency, $f\approx 33$ Hz for electrons or $f\approx 18$ mHz for protons, corresponds to zero kinetic energy. Global intuition is made difficult due to the non-linear relation between energy and frequency differential elements. Finally, one should keep in mind that 
if the frequency of a given GW is obviously proportional to its energy, 
in the specific case of a charged particle orbiting in a magnetic field, frequency is inversely proportional to the energy {\it of the emitting particle}.

\section{GW power and density from galactic cosmic-rays}

\begin{figure}
    \centering
\includegraphics[width=0.5\textwidth]{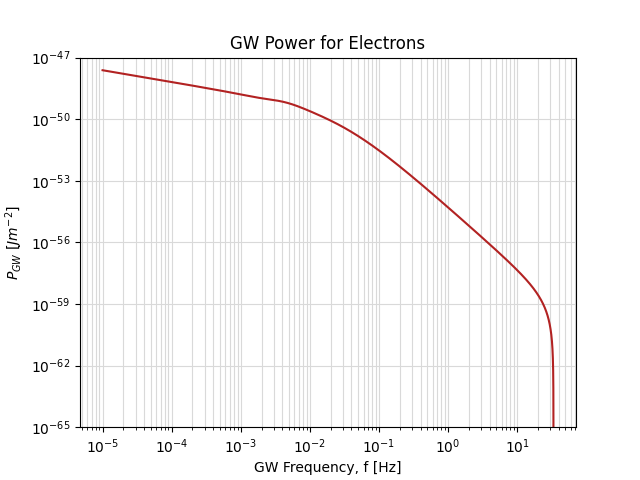}
    \caption{GW power received on Earth on a unit surface from the distribution of galactic electrons.}%
    \label{elect_power}
\end{figure}

\begin{figure}
    \centering
\includegraphics[width=0.5\textwidth]{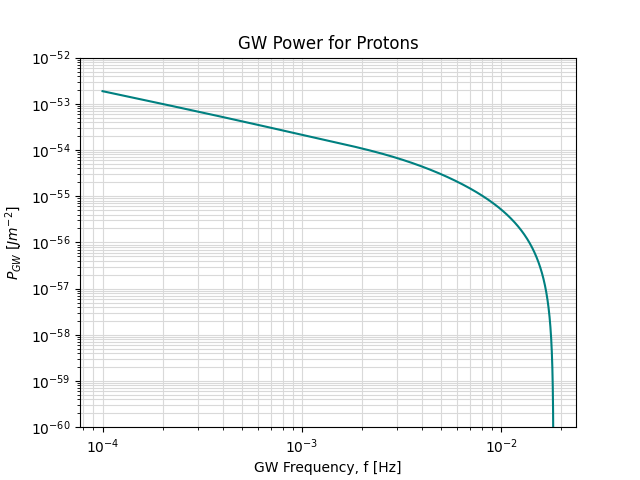}
    \caption{GW power received on Earth on a unit surface from the distribution of galactic protons.}%
    \label{prot_power}
\end{figure}

The differential flux of GW power on Earth can now be obtained by performing the convolution of the emitted power per particle with the spectrum of CR integrated over the distance $R$:
\begin{equation}
    \frac{dP}{dSdf}=\int_0^{R_{max}}\frac{d^2N}{dVdf}(f)P(f)dR.
\end{equation}
This expression for the flux of GW power passing through a unit surface detector on Earth is an approximation in at least two ways. First, it assumes the emission to be isotropic, which is not true but remains statistically relevant. Second, it disregards the specific position of the Earth within the Galaxy and the detailed shape of the latter. It is however easy to check that the weak (linear) dependence of the result with distance makes this assumption precise enough for this first estimate. The resulting spectra are given in Figs.(\ref{elect_power}) and (\ref{prot_power}). As expected, the spectrum falls down to zero at the critical frequency corresponding to a vanishing kinetic energy. The overall shape is, however, non-trivial and results from many different competing effects. The highest power flux is reached at low frequencies (although the density of cosmic-rays is very low). The maximum energy of galactic cosmic-rays is around $10^{15}$ eV, which corresponds to $10^{-8}$ Hz. The associated power flux is $P_{\text{max}}^{\text{elec}}\approx 10^{-47}$ J.m$^{-2}$ for electrons and $P_{\text{max}}^{\text{prot}}\approx 10^{-50}$ J.m$^{-2}$ for protons.\\

\begin{figure}
    \centering
\includegraphics[width=0.5\textwidth]{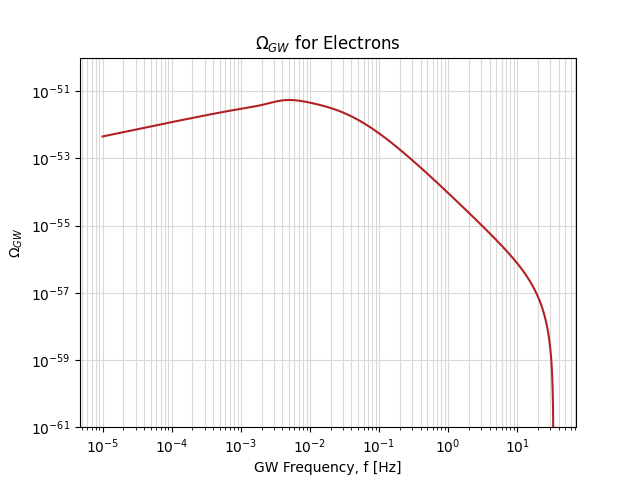}
    \caption{Energy density of GWs from galactic cosmic electron per logarithmic frequency interval, normalized to the critical density.}%
    \label{fig_omega_elec}
\end{figure}

\begin{figure}
    \centering
\includegraphics[width=0.5\textwidth]{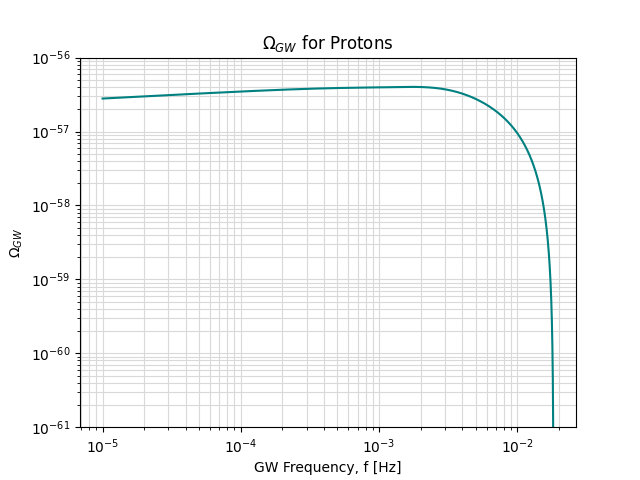}
    \caption{Energy density of GWs from galactic cosmic protons per logarithmic frequency interval, normalized to the critical density.}%
    \label{fig_omega_prot}
\end{figure}

It is usual to define the normalized energy density of GWs per logarithmic frequency interval,
\begin{equation}
    \Omega(f)=\frac{f}{\rho_c}\frac{d\rho}{df},
\end{equation}

where $\rho_c=3H_0^2/(8\pi G)$ is the critical energy density of the universe, $H_0$ is the Hubble parameter today, and $\rho$ in the energy density of GWs. The resulting functions are displayed in Fig.(\ref{fig_omega_elec}) for electrons and Fig.(\ref{fig_omega_prot}) for protons. Interestingly, $\Omega(f)$ exhibits a maximum, respectively at $f_{\text{max}}^{\text{elec}}\approx 5$ mHz for electrons, and $f_{\text{max}}^{\text{prot}}\approx 2$ mHz for protons. Quite surprisingly, this falls within the LISA interferometer \cite{Baker:2019nia} band-pass (which anyway has no experimental consequence as the amplitude of the cosmic-ray signal is way below the sensitivity of the instrument). It simply happens to be the case that the orbital frequency of supermassive black holes close to coalescence is of the same order of magnitude than the Larmor pulsation of a cosmic-ray spiraling in the galactic magnetic field.\\

The shape and the peaking frequency (right in the LISA band) of $\Omega(f)$ are non-trivial and specific to this process. The value is, as expected, very small and way below anything that could be measured in a reasonable future. We stress once more that we are not here testing an exotic theory. Should it be the case, it would be interesting only if the prediction was actually measurable. Just the other way around, we are using 
well-known physics to calculate an elegant and existing physical process. 

\section{Extra-galactic component}

The term ``extra-galactic cosmic-rays" usually refers to particle moving between galaxies. For the study carried out here, it is obvious that the associated signal would be completely negligible -- at least for $\Omega(f)$ -- because of both the very small fluxes and the negligible magnetic fields. 

Still, there are cosmic-rays diffusing within other galaxies. They are not in the intergalactic space but they are extra-galactic from our point of view. Although they are completely irrelevant for any direct measurement, they do emit GWs that reach the Earth. They are the important extra-galactic component for this study.\\

As a very crude -- but sufficient for this work -- approximation we shall assume that our galaxy is typical. The received power then reads

\begin{equation}
    \frac{dP_{extra-gal}}{dfdS}=\int_0^{R_{max}}\frac{dP^{full}_{gal}}{df}(R)\frac{dn}{dV}dR,
    \label{extrag}
\end{equation}
where $dn/dV$ is the mean number of galaxies per unit volume, and
\begin{equation}
    \frac{dP^{full}_{gal}}{df}=\int_0^{R_{gal}}\frac{d^2N}{dVdf}P(f)4\pi R^2dR.
    \label{extrag2}
\end{equation}
The sensitivity of the result to the precise value of $R_{max}$ was checked to be very weak and the plots presented here were obtained for one half of the Hubble radius.
Several remarks are in order. First, one should keep in mind that $dP^{full}_{gal}/df$ is {\it not} what was previously calculated but is the GW signal associated with a full galaxy considered as a source. Second, it is worth noticing that in Eq. (\ref{extrag}), the frequency dependence in {\it not}, as previously, in the source spectrum which is convoluted with the power but in the (differential) power itself. Third, and most importantly, the $R-$dependence of the integrand in Eq. (\ref{extrag}) is because in Eq. (\ref{extrag2}), the frequency must be red-shifted (the red-shift $z$ being, itself, a function of $R$). The isotropy of the distribution of galaxies at large scale ensures that the signal should be mostly isotropic. Small fluctuations -- whose amplitude  depends upon the assumed angular resolution -- are of course expected but are not of major significance for this work.\\

\begin{figure}
    \centering
\includegraphics[width=0.5\textwidth]{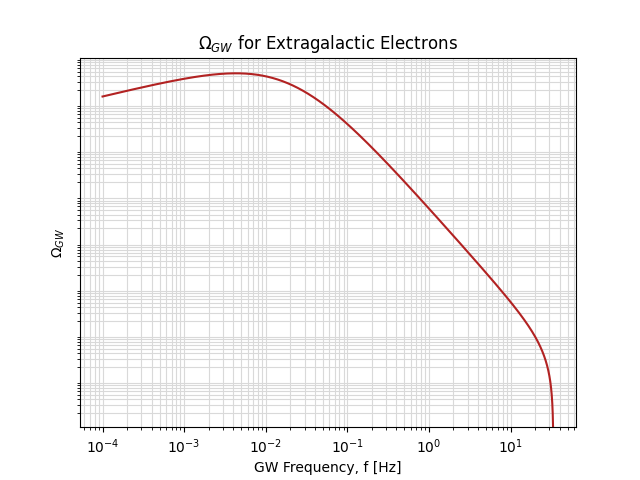}
    \caption{Shape of the energy density of GWs from extra-galactic cosmic electron per logarithmic frequency interval. Arbitrary normalization.}%
    \label{fig_omega_elec_extra}
\end{figure}

\begin{figure}
    \centering
\includegraphics[width=0.5\textwidth]{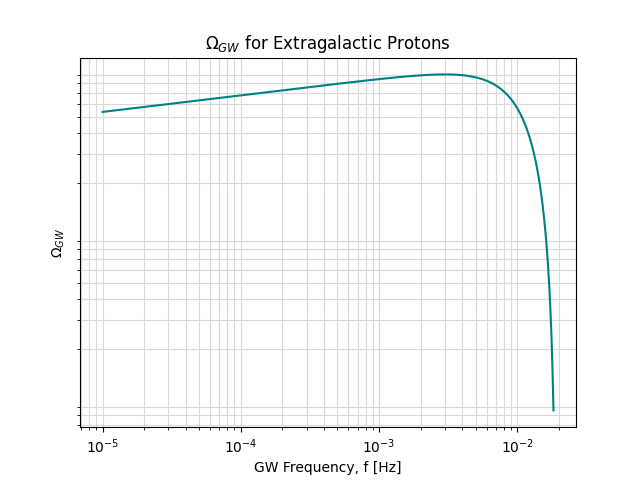}
    \caption{Shape of the energy density of GWs from extra-galactic cosmic protons per logarithmic frequency interval. Arbitrary normalization.}%
    \label{fig_omega_prot_extra}
\end{figure}

The resulting $\Omega(f)$ spectra are given in Figs. \ref{fig_omega_elec_extra} and \ref{fig_omega_prot_extra} for electrons and protons respectively. The "smoothing" due to the redshift can clearly be seen around the cutoff frequency. We have only displayed the shape as the modelling it too rudimentary to get a reliable estimate of the amplitude which is, anyway, extremely small and not relevant in itself. 

\section{Conclusion}

In this brief article, we have first underlined some surprising features of the emission of gravitational waves from a charged particle in a magnetic field. In particular, we have shown that the generated strain depends neither on the amplitude of the magnetic field, nor on the energy of the particle (in the relativistic limit). We have then carefully investigated the radiated power and have suggested a phenomenological expression, approximately valid in all regimes. The comparison with the electromagnetic synchroton power shows that, for hypothetical ultra-heavy particles, the GW power could dominate. The formalism was applied to the distribution of cosmic electrons and protons -- whose spectrum had to be re-written as a function of the gyration frequency. The resulting stochastic background of GWs has a maximum around a few mHz for the galactic component. The shape of $\Omega(f)$ is non-trivial and was calculated for both the galactic and extra-galactic components. The integrated red-shift effect changes the shape of the extra-galactic component near the cutoff frequency.\\

It should be noticed that the approximation of constant magnetic field is obviously an over-simplification as the interstellar magnetic field is highly stochastic, varying in direction over tens of parsecs, so the gravitational wave emission may fluctuate highly over the whole galaxy. The RMS magnitude of the associated power might depend on the coarse graining scale. This is something that should be investigated in future works.\\

The frequency spectrum of the gravitational waves emitted by a relativistic particle orbiting around a magnetic field we used in this work, presented in Eq. (\ref{GW power spectrum}) of section III, is in direct agreement with \cite{Chen:2021rvg}, and in line with the results given in \cite{Davis:1972dm,Chitre:1972fv,PhysRevD.7.1002,Ternov:1975ut}. It is however not the only viewpoint on the question. Another view has also been proposed in \cite{KHALILOV197243, SOKOLOV19781} and could be investigated in a future work.\\

In parallel we have also tried to make a little epistemic point which, we believe, is meaningful beyond this specific work. Whereas predictions are interesting only if they can be experimentally checked when trying to test a new theory or an exotic model, this is {\it not} true when dealing with the application of well established theories within their known domain of validity. Physics is not just about discovering the fundamental laws of Nature, it is also -- assuming those laws are (approximately) known -- about trying to discover what is actually happening in the Universe. In that case, the fact that we focus on a phenomenon extremely unlikely to be observed -- but still intriguing and elegant -- is precisely what makes the investigation relevant.\\

\section{Aknowledgements}

We would like to thank David Maurin for very helpful discussions.

\bibliography{refs.bib}

 \end{document}